\ifx\justbeingincluded\undefined
\documentclass{styles/svproc}
\usepackage[utf8]{inputenc}
\usepackage{latexsym}
\usepackage{amsmath}
\usepackage{amssymb}
\usepackage{graphicx}
\usepackage{cite}
\usepackage{url}

\newcommand{\OO}{\mathcal{O}}
\DeclareMathOperator{\dx}{d\mathit{x}}
\DeclareMathOperator{\Per}{Per}
\providecommand{\brac}[1]{\left(#1\right)}
\providecommand{\TV}[1]{\mathrm{TV}[#1]}
\DeclareMathOperator*{\argmin}{argmin}
\newcommand{\setof}[2]{\left\{{#1}\,:\,{#2}\right\}}
\providecommand{\abs}[1]{\left\lvert#1\right\rvert}

\begin{document}
\mainmatter              % start of a contribution
\fi

\title{Integrated modeling and validation for phase change with natural convection}
\titlerunning{Phase Change with Convection}  % abbreviated title (for running head)
\author{Kai Sch\"uller\inst{1,2} \and Benjamin Berkels\inst{1} \and Julia Kowalski\inst{1}}
\authorrunning{K. Sch\"uller, B. Berkels, J. Kowalski} % abbreviated author list (for running head)
\tocauthor{Kai Sch\"uller, Julia Kowalski, Benjamin Berkels}
\index{Sch\"uller, K.}
\index{Kowalski, J.}
\index{Berkels, B.}
\institute{AICES Graduate School, RWTH Aachen University, Schinkelstr. 2, 52062 Aachen, Germany.\\
\and
\email{schueller@aices.rwth-aachen.de}}

\maketitle              % typeset the title of the contribution

\begin{abstract}
Water-ice systems undergoing melting develop complex \linebreak spatio-temporal interface dynamics and a non-trivial temperature field.
In this contribution, we present computational aspects of a recently conducted validation study that aims at investigating the role of natural convection for cryo-interface dynamics of water-ice.
We will present a fixed grid model known as the enthalpy porosity method \cite{brent1988enthalpy,kumar2017influence}.
It is based on introducing a phase field and employs mixture theory.
The resulting PDEs are solved using a finite volume discretization.
The second part is devoted to experiments that have been conducted for model validation. The evolving water-ice interface is tracked based on optical images that shows both the water and the ice phase.
To segment the phases, we use a binary Mumford Shah method, which yields a piece-wise constant approximation of the imaging data. Its jump set is the reconstruction of the measured phase interface.
Our combined simulation and segmentation effort finally enables us to compare the modeled and measured phase interfaces continuously. We conclude with a discussion of our findings.
% We would like to encourage you to list your keywords within
% the abstract section using the \keywords{...} command.
\keywords{phase change, finite volume method, OpenFOAM, image segmentation}
\end{abstract}

\section*{Nomenclature}
\begin{tabular}{p{0.1\linewidth}p{0.4\linewidth}}
	$A$ & Kozeny-Carman relation  \\
	$c_p$ & heat capacity \\
	$\bar{c}_p$ & averaged heat capacity \\
	$c_1,c_2$  & gray scale values \\
	$C$ & mushy zone constant \\
	$f$ & phase mass fraction  \\ 
	$\mathbf{F}$  & phase interaction force \\
	$\mathbf{g}$ & gravitational acceleration \\
	$h$ & enthalpy \\
	$h_m$ & latent heat of melting \\
	$k$ & thermal conductivity \\
	$p$ & pressure \\
	$\mathbf{S}$ & Boussinesq term \\
	$T$ & temperature  \\ 
\end{tabular} 
\begin{tabular}{p{0.1\linewidth}p{0.4\linewidth}}
$T_S$ & solidus temperature  \\
$T_L$ & liquidus temperature   \\ 
$T_m$ & melting temperature  \\ 
$T_\text{init}$ & initial PCM temperature   \\
$T_w$ & wall temperature  \\
$\mathbf{u}$ & velocity field \\
$V$ & volume \\
$\gamma$& phase volume fraction \\
$\epsilon$ & small constant \\
$\eta$ & dynamic viscosity \\
$\Theta$ & temperature deviation \newline($\Theta=T-T_m$) \\
$\rho$ & density \\
$\bar{\rho}$ & partial density \\
\end{tabular} 

\section{Introduction}
Phase change processes play an important role in a variety of present-day research fields and industrial applications.
A material that undergoes phase change, a so-called phase change material (PCM), absorbs and releases heat at a constant temperature $T_m$ or within a certain phase change temperature range, bounded by the liquidus temperature $T_L$ and the solidus temperature $T_S$.
PCMs are particularly relevant to thermal energy storage (TES) systems, because of their large storage density compared to non-latent TES systems (5 - 14 times more heat per unit volume than sensible storage materials \cite{sharma2009review}).
A TES system is an attractive technology because it is the most appropriate method to correct the gap between demand and supply of energy \cite{akeiber2014review}.
This becomes very important in the context of renewable energy sources, because most of them depend on time-varying environmental parameters, such as the wind speed (for wind power plants) or the duration of solar irradiation (for solar power plants).
TES systems are also used for cooling applications, e.g. to protect electrical devices. The cheapest PCM for cooling applications is water-ice. Its melting temperature is 0\,$^\circ$C. Beyond this industrial application, the process of water-ice melting can be found in a variety of scientific areas, e.g. glaciology or ice sheet modeling.

To simulate phase change heat transfer both the sensible and the latent heat release or storage must be considered, which translates into a moving boundary problem as the interface might propagate or retrieve.
Such problems can be solved either with fixed- or deforming grid methods, or a combination of both \cite{voller1997overview}.
Even though deforming grid methods are in general more accurate than fixed grid methods in terms of localizing the phase interface, fixed-grid methods are computationally much more efficient.
The major advantage of fixed grid methods is that the numerical treatment of the phase change can be achieved through simple modifications of existing numerical methods, which allows to model phase change for a variety of complex phase change systems with relative ease \cite{voller1990fixed}.
When the liquid phase of the PCM is convecting, the fluid flow can have a considerable impact on the heat transfer within the system.
Therefore, it is necessary to both solve for the heat transfer and the fluid flow.
A popular method that is used for such phase change processes is the so-called enthalpy porosity method \cite{brent1988enthalpy,kumar2017influence}.

Unfortunately, there exists no analytical solution to verify phase change models with natural convection.
However, one-dimensional phase change without natural convection can be verified by comparison to the analytical solution of the Stefan problem, which has been already done with great success for the enthalpy-porosity method \cite{konig2017comprehensive}.
To validate phase change with natural convection, experiments must be used.
A very common benchmark is the melting of a PCM, which is driven by an isothermal vertical wall in a rectangular cavity.
The majority of these experiments include PCMs with a melting temperature higher than $0\,^\circ$C.
Examples are gallium \cite{gau1986melting} and n-octadecane \cite{ho1984heat}.
Similar experiments exist for water-ice \cite{schutz1992melting}.

In this contribution, we present a fixed grid model that uses the enthalpy porosity method to simulate phase change with natural convection.
In order to validate the model, we conducted our own experiments, which are similar to existing benchmark tests but with high spatio-temporal resolution.
The data consist of optical images that show the motion of the phase interface over time.
To extract the phase interface from the optical images, we use binary Mumford-Shah segmentation.
This allows for a quantitative comparison between the model and the experimental results.

\section{Model}
\subsection{Physical situation}
\begin{figure}
	\centering
	\includegraphics[width=0.7\linewidth]{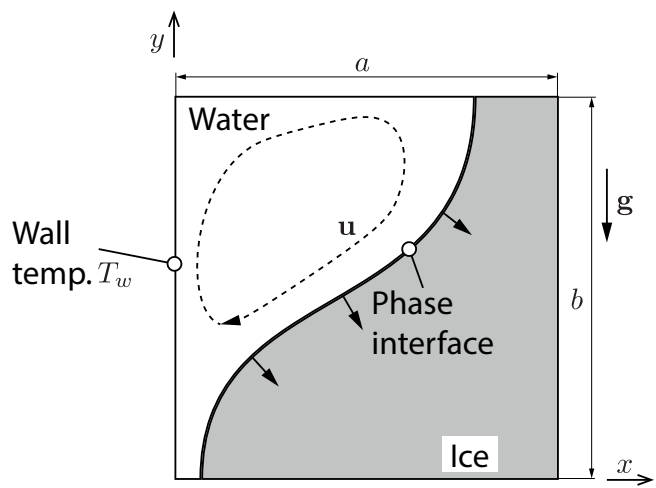}
	\caption{Schematic of the physical situation. A two-dimensional cavity is filled with a PCM, which is present both in liquid, as well as its solid phase. Both phases are separated by the phase interface. The left boundary is held at constant temperature $T_w$.}
	\label{fig:experimentproblem}
\end{figure}

The physical situation is sketched in figure \ref{fig:experimentproblem}.
A two-dimensional cavity of size $a\times b$ is filled with an initially solid phase change material (PCM) of temperature $T_\text{init}$.
Due to an imposed temperature $T_w$ at the left boundary, which is higher than the melting temperature of the PCM $T_m$, the PCM heats up locally and changes its phase from solid to liquid.
Both phases are separated by a phase interface.
The shape of the phase interface is mainly defined by natural convection.
In the presence of gravitational acceleration $\mathbf{g}$, the density variation in the liquid phase induces natural convection, which manifests as a clockwise rotational flow field $\mathbf{u}$ within the liquid phase.
The presented approach is applicable to a variety of PCMs, e.g. metals or waxes. In this study, we will however focus on water-ice.

\subsection{Model equations}
To formulate a fixed-grid mathematical model that describes the physical problem of phase change with natural convection, either volume-averaging or classical mixture theory can be utilized.
Here, we will shortly sketch the latter approach based on mixture theory.
Interested readers can find a comprehensive derivation of the mixture equations using volume-averaging in \cite{ni1991volume}.
The basic idea of mixture theory is to introduce a scalar field, which stores the information of the PCM state. Here, we use the phase volume fraction, which is defined as
\begin{align}
	\label{eq:phaseVolumeFractionGeneral}
	\gamma_i=\frac{V_i}{\sum\limits_i V_i}
\end{align}
in which $V_i$ is the volume of phase $i$ in a control volume.
From equation \eqref{eq:phaseVolumeFractionGeneral}, it can be seen that the value of the phase volume fraction is always between zero and unity. We further assume full saturation, i.e.
\begin{equation}
	\label{eq:volumeFractionSaturation}
	\sum\limits_i \gamma_i=1
\end{equation}
The partial density of phase $i$ is then given by
\begin{equation}
	\label{eq:partialDensityGeneral}
	\bar{\rho}_i=\gamma_i\rho_i
\end{equation}
in which $\rho_i$ is the density of phase $i$. The mass fraction of phase $i$ is
\begin{equation}
\label{eq:massFraction}
f_i=\frac{\bar{\rho_i}}{\sum\limits_i \bar{\rho}_i}
\end{equation}
In this work, we are interested in a two-phase system, which is given by a solid and a liquid phase.
Substituting equations \eqref{eq:volumeFractionSaturation} and \eqref{eq:partialDensityGeneral} into the mass fraction \eqref{eq:massFraction} yields an explicit relation for the liquid phase
\begin{equation}
	\label{eq:massFractionForLiquidPhase}
	f_L=\frac{\gamma_L\rho_L}{\gamma_L\rho_L+\gamma_S\rho_S}
\end{equation}
In the special case of $\rho_L=\rho_S$, equation \eqref{eq:massFractionForLiquidPhase} reduces to $f_L=\gamma_L$ and analogously $f_S=\gamma_S$.

According to \cite{bennon1987continuum}, the three mixture balance laws are obtained by summing the balance laws for the individual phases, i.e. conservation of mass, momentum and energy for the liquid and solid phase.
After some simplifications and introducing a set of mixture variables and parameters, we derive a system that accounts for incompressible mixture flow and phase change coupled to natural convection. It is given by
\begin{align}
	\label{eq:massEquation1}
	\nabla\cdot \mathbf{u}&=0\\
	\label{eq:momentumEquation1}
	\frac{\partial \left(\rho \mathbf{u}\right)}{\partial t}+\nabla\cdot\left( \rho \mathbf{u}\otimes\mathbf{u} \right)&=-\nabla p +\nabla\cdot \left( \eta \nabla \mathbf{u} \right)+\mathbf{F}+\mathbf{S}(T)\\
	\frac{\partial \left( \rho h \right)}{\partial t}+\nabla\cdot \left( \rho \mathbf{u} h \right)&=\nabla\cdot \left( k\nabla T \right)
\end{align}
in which
\begin{align}
	\rho&=\gamma_S\rho_S+\gamma_L\rho_L\\
	k&=\gamma_S k_S+\gamma_L k_L\\
	\label{eq:mixtureEnthalpyMassFraction}
	h&=f_S h_S+f_L h_L\\
	\eta&=\eta_L\\
	\mathbf{u}&=\mathbf{u}_S=\mathbf{u}_L
\end{align}
are the mixture density $\rho$, mixture thermal conductivity $k$, mixture enthalpy $h$, mixture dynamic viscosity $\eta$ and mixture velocity $\mathbf{u}$.
Note that in the local presence of both phases, we assume them to move at the same velocity.
This assumption is appropriate, because relative phase motion can be neglected.

Equation \eqref{eq:momentumEquation1} is the conservation of momentum, which includes two additional terms, namely a temperature dependent Boussinesq approximation term $\mathbf{S}(T)$ and a phase interaction force term $\mathbf{F}$.
The Boussinesq approximation term accounts for free convection due to buoyancy and is defined as
\begin{equation}
	\mathbf{S}(T)=\mathbf{g}\rho(T)
\end{equation}
in which $\rho(T)$ is a polynomial fit to tabulated density data. It should be noted that the Boussinesq approximation is only valid if the density variation is small, which is a valid assumption for water.

The phase interaction force $\mathbf{F}$ accounts for momentum production due to phase interactions \cite{bennon1987continuum}. According to \cite{voller1987fixed}, the flow regime within cells that contain portions of both phases can be interpreted as a porous medium. Hence, the flow can be described by Darcy's law. This behavior can be accounted for by defining
\begin{align}
	\mathbf{F}=A\mathbf{u}
\end{align}
$A$ is large in the liquid phase ($\gamma_L=1$) and small in the solid phase ($\gamma_L=0$). This allows for flow in the liquid phase, whereas it suppresses it in the solid phase. A commonly used continuous function with this properties is the Kozeny-Carman relation \cite{voller1987fixed}
\begin{equation}
	\label{eq:darcyTerm}
	A=-C\frac{\left( 1-\gamma_L \right)^2}{\gamma_L^3+\epsilon}
\end{equation}
Here, $\epsilon$ (typically $\epsilon=10^{-6}$) is a stabilizing parameter that is used in order to prevent division by zero and $C$ denotes the mushy zone constant.
It should be noted that $C$ has no direct physical significance and has to be calibrated with data.
In non-isothermal phase change processes the PCM develops a mushy region rather than a sharp phase interface.
In this case, adjusting the mushy zone constant can be exploited to model the resulting porosity near the mushy phase interface.

\subsection{Source-based method for phase change}
In order to solve the energy equation, which is given in enthalpy form, we need to introduce an equation that relates the enthalpy to the temperature.
The enthalpies of the solid and liquid phases are given by
\begin{align}
	\label{eq:enthalpySolid0}
	h_S&=\int_{T_m}^T c_{p,S}\text{d}T\\
	\label{eq:enthalpyLiquid0}
	h_L&=\int_{T_m}^T c_{p,L}\text{d}T+h_m
\end{align}
in which $h_m$ is the latent heat of melting and $c_{p,S}$ as well as $c_{p,L}$ are the heat capacities of the solid and liquid phase, respectively.

Over a temperature range of 20\,K, the percentage heat capacity change is in the order of 5\,\% for ice and 1\,\% for water. If we assume phase-wise constant heat capacities, which is a valid approximation as long as the temperature range within the PCM is small, equations \eqref{eq:enthalpySolid0} and \eqref{eq:enthalpyLiquid0} simplify to
\begin{align}
	\label{eq:enthalpySolid}
	h_S&=\bar{c}_{p,S}(T-T_m)\\
	\label{eq:enthalpyLiquid}
h_L&= \bar{c}_{p,L}(T-T_m)+h_m
\end{align}
From equation \eqref{eq:massFractionForLiquidPhase}, it can be seen that if the densities of the solid and liquid phases are equal, the mass fraction \eqref{eq:massFraction} has the same value as the volume fraction, i.e. $f_k=\gamma_k$.
Within the scope of this work, we will restrict ourselves to this situation and substitute the mass fraction in the mixture enthalpy equation \eqref{eq:mixtureEnthalpyMassFraction} by the volume fraction, which yields
\begin{equation}
\label{eq:mixtureEnthalpyVolumeFraction}
h=\gamma_S h_S+\gamma_L h_L
\end{equation}
Substituting the approximations for the solid \eqref{eq:enthalpySolid} and liquid enthalpy \eqref{eq:enthalpyLiquid} into the equation for the mixture enthalpy \eqref{eq:mixtureEnthalpyVolumeFraction} yields
\begin{equation}
	\label{eq:finalMixtureEnthalpy}
	h=\bar{c}_p (T-T_m)+\gamma_Lh_m
\end{equation}
in which $\bar{c}_p=\gamma_L \bar{c}_{p,L} + \gamma_S \bar{c}_{p,S}$ is the mixture heat capacity.
We can now substitute the mixture enthalpy \eqref{eq:finalMixtureEnthalpy} into the energy equation, which yields
\begin{equation}
	\label{eq:finalEnergyEquation}
	\frac{\partial(\rho \bar{c}_p \Theta)}{\partial t}+\nabla\cdot \left( \rho \mathbf{u} \bar{c}_p \Theta \right)=\nabla\cdot \left( k\nabla \Theta \right)-h_m\left( \frac{\partial \left( \rho \gamma_L \right)}{\partial t}+\nabla\cdot \left( \rho \mathbf{u}\gamma_L \right) \right)
\end{equation}
in which $\Theta=T-T_m$ denotes the deviation from the melting temperature.

The left-hand side and the first term of the right-hand side of equation \eqref{eq:finalEnergyEquation} matches the standard transient convection-diffusion energy equation that describes sensible heat transfer.
The remaining term accounts for the latent heat transfer due to phase change.

\subsection{Solution algorithm for the energy equation}
Equation \eqref{eq:finalEnergyEquation} contains two unknowns, namely the temperature $\Theta$ and the liquid volume fraction $\gamma_L$. These two fields, however, are intrinsically coupled. In order to solve equation \eqref{eq:finalEnergyEquation}, a relation between the temperature and the liquid volume fraction is required.
In our work we follow \cite{rosler2014modellierung} and use a piecewise linear function
\begin{align}
\label{eq:gammaAsFuncOfT}
\gamma_L= \begin{cases}
0, &T<T_S\\
\frac{T-T_S}{T_L-T_S}, &T_S \leq T \leq T_L\\
1, &T>T_L
\end{cases}
\end{align}
This approach assumes that the phase change occurs within a narrow temperature range $T_L-T_S$, rather than at a fixed temperature.

Following \cite{voller1987fixed}, we linearize equation \eqref{eq:finalEnergyEquation} and introduce an iterative corrector approach
\begin{equation}
	\label{eq:finalEnergyEquationLinearized}
	\frac{\partial(\rho c_p \Theta^{k+1})}{\partial t}+\nabla\cdot \left( \rho \mathbf{u} c_p \Theta^{k+1} \right)=\nabla\cdot \left( k\nabla \Theta^{k+1} \right)-h_m\left( \frac{\partial \left( \rho \gamma_L^k \right)}{\partial t}+\nabla\cdot \left( \rho \mathbf{u}\gamma_L^k \right) \right)
\end{equation}
in which $\gamma_L^k$ is the known volume fraction of the previous iteration $k$ and $\Theta^{k+1}$ is the solution variable of the current iteration.
The updated temperature $\Theta^{k+1}$ does not match the temperature determined through relation \eqref{eq:gammaAsFuncOfT} based on the volume fraction of the previous iteration $k$.
Therefore, an energy conserving updating of the volume fraction is used \cite{konig2017comprehensive,faden2018implicit}
\begin{equation}
	\label{eq:liquidFractionUpdateEquation}
	\gamma_L^{k+1}=\max\left[\min\left[\gamma_L^k+\lambda \frac{c_p}{h_m}\left( \Theta^{k+1}-\Theta_\text{cons}^{k+1} \right),1\right],0\right]
\end{equation}
with
\begin{equation}
	\label{eq:consistentTemperature}
	\Theta_\text{cons}^{k+1}=T_S+\left( T_L-T_S \right)\gamma_L^k-T_m
\end{equation}
in which $\lambda$ is a relaxation factor.
According to \cite{voller1990fixed}, values between 0.5 and 0.7 provide efficient convergence for both one- and two-dimensional problems.
The consistent temperature equation \eqref{eq:consistentTemperature} directly follows from the volume fraction temperature relation \eqref{eq:gammaAsFuncOfT}.
Equation \eqref{eq:liquidFractionUpdateEquation} further assures that no over- and undershooting of the volume fraction occurs, i.e. the values will be always between zero and unity. 

\subsection{Summary of the iterative solution procedure}
Incorporating the stated equations, the following iterative solution procedure is applied to solve the energy equation \eqref{eq:finalEnergyEquation}:
\begin{enumerate}
	\item Either set an initial liquid volume fraction $\gamma_L^k$ if it is the first time step or use the volume fraction of the previous time step.
	\item Solve the linearized energy equation \eqref{eq:finalEnergyEquationLinearized} for $\Theta^{k+1}$.
	\item Calculate the temperature $\Theta_\text{cons}^{k+1}$, which is consistent to the volume fraction from the previous iteration using equation \eqref{eq:consistentTemperature}.
	\item Update the volume fraction $\gamma_L^{k+1}$ using equation \eqref{eq:liquidFractionUpdateEquation}. Go back to step 2 if the convergence threshold is not reached, i.e. if the error of the volume fraction is not smaller than a certain tolerance.
\end{enumerate}

\subsection{Implementation}
For this work we used OpenFOAM, which is an object oriented open source C++ library to solve PDEs \cite{weller1998tensorial,jasak2007openfoam}.
We implemented the enthalpy porosity method by extending \textit{buoyantBoussinesqPimpleFoam} (OpenFOAM 5.0), which is a transient solver for buoyant, turbulent flow of incompressible fluids that uses the PIMPLE algorithm for pressure velocity coupling.

\subsection{Mushy zone constant sensitivity}
\begin{figure}
	\centering
	\includegraphics[width=.9\linewidth]{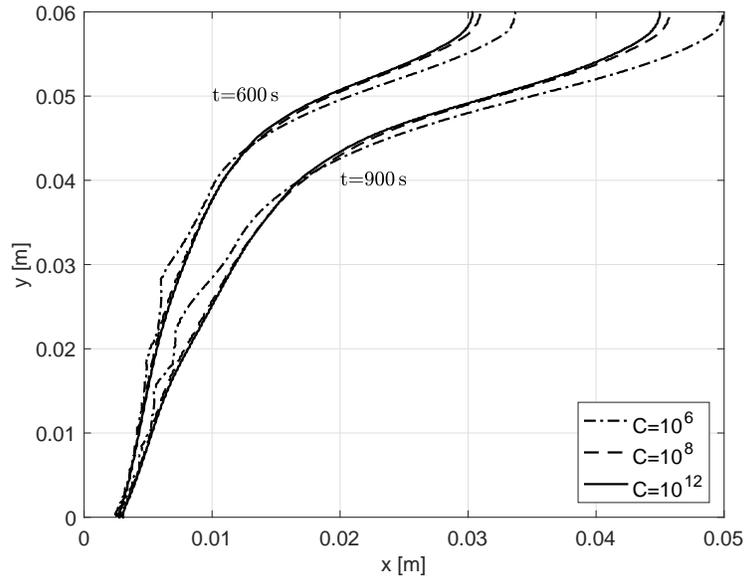}
	\caption{Melting from an isothermal vertical wall using different mushy zone constants.}
	\label{fig:mushyzoneconstantcomparison}
\end{figure}

The  sensitivity of the mushy zone constant with respect to the resulting phase interface has been studied for gallium \cite{kumar2017influence} and for lauric acid \cite{kheirabadi2015effect}.
Both studies conclude that the mushy zone constant significantly influences the shape of the resulting phase interface.
So it should be chosen carefully in order to obtain reasonable results.
To our knowledge, such sensitivity studies have not been conducted for water-ice PCMs.
Since we want to validate our phase change simulations against water-ice PCMs, we also studied the results for different mushy zone constants.
Figure \ref{fig:mushyzoneconstantcomparison} shows the phase interfaces at 600 and 900 seconds.
The simulations have been conducted on a quadratic uniform mesh of 102,400 quadrilateral cells using adiabatic boundaries, except for the left boundary at which a Dirichlet condition of $30.5\,^\circ$C is applied.
Furthermore, we use no-slip conditions at all boundaries and temperature dependent thermophysical material properties.
The initial temperature is $T_{init}=-20\,^\circ$C.
It can be seen, that the phase interface oscillates for a mushy zone constant of $C=10^6$. Increasing the mushy zone constant yields a smoother phase interface. Furthermore, the plot shows that the phase interface converges if the mushy zone constant is increased. Based on our findings, we chose a value of $C=10^{10}$ for all following simulations.

\section{Validation experiment}
\subsection{Apparatus and instrumentation}
\begin{figure}
	\centering
	\includegraphics[width=1\linewidth]{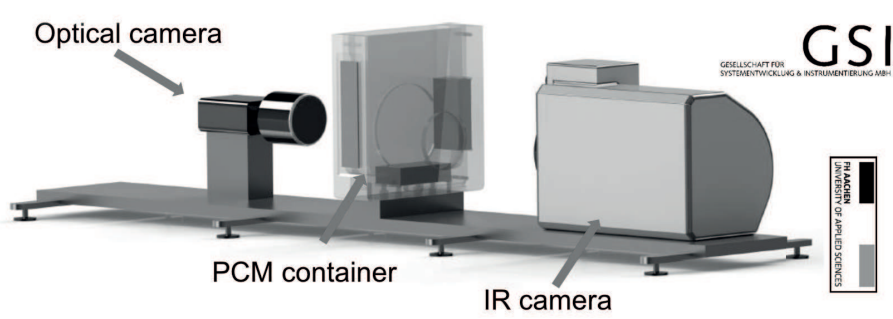}
	\caption{Experiment assembly.}
	\label{fig:experimentcad}
\end{figure}

The experimental setup is shown in figure \ref{fig:experimentcad}.
It consists of a PCM container, which is made out of plexiglas and one optical as well as one infrared camera.
In order to observe the melting process, the PCM container has two circular windows of different materials, namely plexiglas for the optical camera and germanium for the infrared camera.
In this work, we will focus on the results of the optical camera.
The PCM container is equipped with two heater blocks of size $30\times60\times20\,\text{mm}^3$ that are in contact with the PCM.
Each heater block contains two heating cartridges, which can be controlled independently.
The heater blocks contain temperature sensors for temperature control.
A feedback loop sustains a predefined temperature by means of a straight-forward feedback control.

For this work, we will study the case of an isothermal vertical wall.
So we do not use the bottom heater.
The dimensions of the inner PCM container are $30\times107\times114\,\text{mm}^3$.

\subsection{Experimental procedure}
Before each experiment, the PCM container was removed from the experiment assembly, which was located in an approximately $22\,^\circ$C warm laboratory.
The container was filled with liquid water and put into a freezer in order to transform the water into ice.
The water level was always above the side heater before freezing.
To obtain better ice qualities, a low power heat source has been placed close to the water surface, so that the solidification proceeds from the bottom to the top of the container.
Otherwise high stresses could have damaged the PCM container.
After a certain amount of time, the ice temperature was approximately -20\,$^\circ$C. Then, the container has been removed from the freezer and reassembled into the experimental setup.
Then, the heater configuration including the target temperature was set.
The logging of data started together with the activation of the vertical heater block. A switch-on temperature of $29\,^\circ$C and a switch-off temperature of $30\,^\circ$C in the feedback control led to temperatures of the heater block oscillating between $28\,^\circ$C and $33\,^\circ$C due to thermal inertia of the heater blocks.

\section{Image segmentation}
To extract the water-ice interface from the optical images, we use the concept of image segmentation. Since there are just two different segments (water and ice), we are facing a two-phase image segmentation problem.

Let $\Omega\subset\mathbb{R}^2$ denote the image plane. Given an image $g:\Omega\rightarrow\mathbb{R}$, we are searching for a piecewise constant segmentation, i.e. two gray values $c_1, c_2$ and a region $\OO\subset\Omega$ that minimizes
\begin{align*}
E[\OO,c_1,c_2]=\int_{\OO} (g-c_1)^2 \,\dx + \int_{\Omega\setminus\OO}(g-c_2)^2\,\dx + \nu \Per(\OO).
\end{align*}
Here, $\Per(\OO)$ is the perimeter of $\OO$, i.e. the length of the phase interface.
For a fixed $\OO$, the optimal gray values $c_1$ and $c_2$ are just the average values of $g$ inside $\OO$ and $\Omega\setminus\OO$ respectively. The minimization with respect to $\OO$ is difficult. Denoting $f_i:=(g-c_i)^2$, we consider the so-called binary Mumford-Shah functional~\cite{MuSh89}
\begin{align*}
E_\textnormal{MS}[\OO]=\int_{\OO} f_1\,\dx + \int_{\Omega\setminus\OO}f_2\,\dx + \nu \Per(\OO).
\end{align*}
Minimizing $E_\textnormal{MS}$ is a nonconvex optimization problem, since the set of subsets of $\Omega$ is not convex. Fortunately, a strongly convex reformulation of this problem is available. The main idea is to replace $\OO$ by a function $w:\Omega\rightarrow\mathbb{R}$. This leads to the functional
\begin{align*}
E_\textnormal{UC}[w]=& \int\limits_{\Omega} w^{2} f_1 \,\dx + \int\limits_{\Omega}\brac{1-w}^{2} f_2 \,\dx + \nu \TV{w},
\end{align*}
where $\TV{w}$ denotes the Total Variation of $w$. Denoting by $\chi_\OO$ the characteristic function of $\OO$, it is easy to show that $E_\textnormal{MS}[\OO]=E_\textnormal{UC}[\chi_\OO]$. In this sense, minimizing $E_\textnormal{UC}$ over $BV(\Omega)$, the set functions with finite Total Variation, is a relaxation of the problem to minimize $E_\textnormal{MS}$ over the subsets of $\Omega$. The former is a strongly convex problem and as such its unique minimizer can be computed efficiently. Moreover, this minimizer encodes a minimizer of the original non-convex problem. One can show \cite{ChDa09,Be09} that
\begin{align*}
w^*=\argmin_{w\in BV(\Omega)}E_\textnormal{UC}[w]\quad\Rightarrow\quad&{\{w^*>0.5\}}\in\argmin\limits_{\OO\subset\Omega}E_\textnormal{MS}[\OO]
\end{align*}
where $\{w>0.5\}$ is the $0.5$-superlevel set of $w$, i.e. $\setof{x}{w(x)>0.5}$.
That means the optimization with respect to $\OO$ can be solved by minimizing $E_\textnormal{UC}$ and thresholding the minimizer. The numerical optimization uses a dual formulation. Recall that the Total Variation is defined as
\[\TV{w}=\sup_{q\in K}\int_\Omega w\nabla\cdot q\dx\]
where $K=\setof{q\in C^\infty_c(\Omega,\mathbb{R}^d)}{\abs{q(x)}\leq1\text{ for all }x\in\Omega}$.
Thus, $E_\textnormal{UC}$ can be minimized by solving a saddle point problem (minimizing in the primal variable $w$, maximizing in dual variable $q$).
Efficient and simple first order algorithms for this are well known~\cite{ChPo11}.

\section{Results and discussion}
\subsection{Image segmentation results}
\begin{figure}
	\centering
	\includegraphics[width=.9\linewidth]{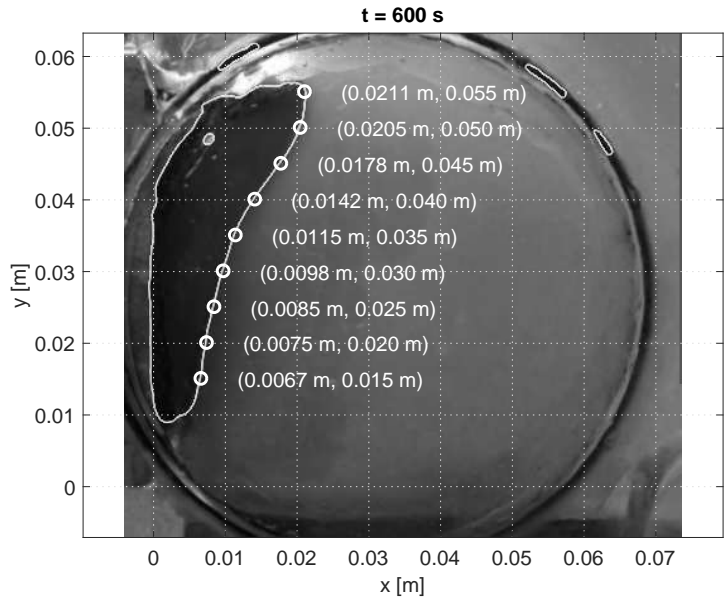}
	\caption{Image taken from the experiment after 600 seconds and segmentation result, as well as the position of nine data points (circles).}
	\label{fig:seg-600}
\end{figure}

\begin{figure}
	\centering
	\includegraphics[width=.9\linewidth]{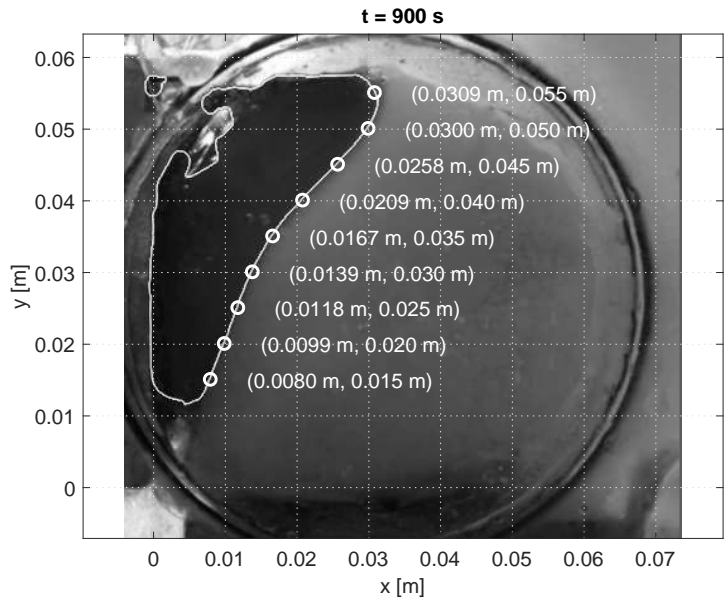}
	\caption{Image taken from the experiment after 900 seconds and segmentation result, as well as the position of nine data points (circles).}
	\label{fig:seg-900}
\end{figure}

Figures \ref{fig:seg-600} and \ref{fig:seg-900} show the images taken from the experiment at 600 and 900 seconds, respectively.
A small portion of the two heater blocks is visible on the left and at the bottom.
The heater blocks have been used as a reference to scale the image pixels to the size of the experiment. It can be seen that the water is dark compared to the ice, which enabled us to apply our segmentation approach.
To compensate for the somewhat non-uniform illumination inherent to our experimental setup, we estimated the background illumination of the scene by applying the morphological opening and closing operator to the first frame of the video and subtracted this illumination estimate from each video frame before applying the segmentation.
Empirically, we found that initial gray values ($c_1$ and $c_2$) of 0.3 and 0.5 work best for the images, which were taken from the experiment. The result of the image segmentation is plotted on top of the images.
Even though the phase interfaces have been detected very good, there are some small artifacts due to similar gray values, e.g. at the circumference of the window in figure \ref{fig:seg-600}.
To better compare the experiment and our numerical results, we arbitrarily chose nine data points (plotted as circles), which are equidistant in $y$-direction.

The effect of natural convection due to buoyancy is clearly visible in the images.
The heating at the left wall causes a decrease in density of the nearby water, which on the other hand induces a flow field in upward direction.
Near the phase interface, heat is absorbed by the ice so that the density increases relative to the average temperature within the water phase.
As the water flows in downward direction along the phase interface, it constantly cools down.
As a consequence the temperature and hence the melting rate is higher near the top compared to the bottom.
The whole process results in a circular flow field in clockwise direction.

\subsection{Comparison with experiment}
In order to compare the experimental results to our model, we used a quadratic uniform mesh of 102,400 quadrilateral cells.
Using a computational domain with the same size of the inner PCM container domain would be computationally inefficient, since most of the space is occupied by ice.
Instead, we use a smaller computational domain of $0.06\times0.06\,$m$^2$, which is large enough to include the entire water phase throughout the simulation.
In order to use temperature dependent material properties for water-ice, we used approximations that fit tabular data from the literature, e.g. from \cite{popiel1998simple}.
To give an example, the density of the water has been approximated using
\begin{equation}
\rho_L=\sum_{i=0}^{3}R_i\left( T-T_{ref} \right)^i+R_4\left( T-T_{ref} \right)^{2.5}
\end{equation}
%\begin{equation}
%	\rho_L=R_0+R_1(T-T_{ref})+R_2(T-T_{ref})^2+R_3(T-T_{ref})^{2.5}+R_4(T-T_{ref})^3
%\end{equation}
in which $T_{ref}=273.15\,$K, $R_0=999.79684\,$kg/m$^3$, $R_1=0.068317355\,$kg/m$^3$/K, $R_2=-0.010740248\,$kg/m$^3$/K$^2$, $R_3=-2.3030988\times10^{-5}\,$kg/m$^3$/K$^3$ and $R_4=0.00082140905\,$kg/m$^3$/K$^{2.5}$.

Except for the left wall at which a Dirichlet condition of $30.5\,^\circ$C is applied, all boundaries are adiabatic.
We further assigned no-slip conditions on all walls.
The initial temperature was set to $T_{init}=-20\,^\circ$C.
The solidus and liquidus temperatures were set to $T_S=-0.05\,^\circ$C and $T_L=0\,^\circ$C, respectively.
The mushy zone constant was set to $C=10^{10}$ based on our findings in the sensitivity analysis.
For the iterative solution of the energy equation, we used a tolerance of $10^{-8}$ for the liquid volume fraction.

\begin{figure}
	\centering
	\includegraphics[width=.9\linewidth]{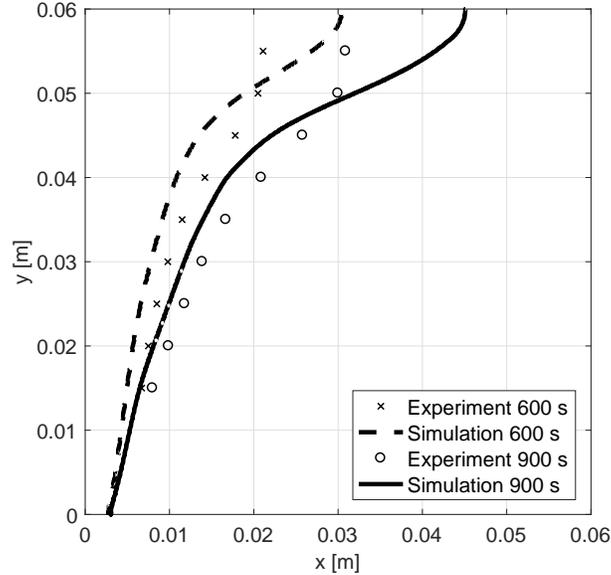}
	\caption{Comparison of the phase interface positions of the experiment and the simulation at 600 and 900 seconds.}
	\label{fig:comparison}
\end{figure}

Figure \ref{fig:comparison} shows the comparison of the phase interface positions of the experiment and the simulation at 600 and 900 seconds.
The phase interface obtained by the simulation qualitatively fits the experimental results, even though there is a significant offset between both results.
It can be seen that the maximum melting rate is located at the top for both the experiment and the simulation results. The maximum error is at the top ($y=0.06\,$m).
It is smaller at 600 seconds, at which the phase interface is captured really well, compared to the results at 900 seconds.
At a height of approximately $0.05\,$m, the phase interface of the simulation and the experiment intersect and below $0.05\,$m the phase interface position of the simulation migrates slower than it was observed in the experiment.

Besides the fact that we simulated the process using a two-dimensional domain instead of a more realistic three-dimensional domain, there are some additional error sources and uncertainties, which could explain the discrepancy between the experiment and the numerical results.
The initial temperature of the ice is inferred from the preparation procedure.
However, we tested the range of possible initial temperatures between $-25\,^\circ$C and $0\,^\circ$C and discovered only a small sensitivity.
A larger error could result from too idealized boundary conditions. We assumed adiabatic walls, except for the boundary at which the heater is located.
In the experiment, the water-ice PCM was in contact to plexiglas walls, which will introduce a heat sink, because the experiment has been conducted in a laboratory with an ambient temperature of $22\,^\circ$C.
The heater has been modeled by using a Dirichlet boundary condition.
However, the heater temperature oscillates between $28\,^\circ$C and $33\,^\circ$C due to the control loop.

\section{Conclusions and outlook}
In this contribution we describe a fixed grid model to simulate phase change processes with natural convection.
The model is based on the enthalpy porosity method, a phase field method, which can be derived from classical mixture theory.
We use an iterative corrector approach to solve the resulting nonlinear energy equation. The final system of PDEs that describes the incompressible mixture flow with phase change has been solved using OpenFOAM.
The method uses a parameter referred to as the mushy zone constant.
A sensitivity study suggests that the mushy zone constant should be high in order to capture the physical regime of water-ice.

In order to validate the model, we conducted experiments in which water-ice was melted from an isothermal vertical wall.
We tracked the water-ice interface using optical images, which resulted in experimental data of the phase interface at high spatio-temporal resolution.
In order to utilize this data, we used the Mumford Shah method to segment the phases in the imaging data and to quantify the phase interface position.
Our results demonstrate the proficiency of this approach for water-ice segmentation in images.
It allows for comparison between the simulation and the experiment.

We observed good qualitative agreement regarding the shape throughout the whole evolution of the phase interface.
Measured from the left boundary, the maximum distance of the phase interface is near the top, which directly follows from the buoyancy-induced flow field in the liquid phase.
Although the results look qualitatively similar, there is, however, a an error between the simulation and experiment in terms of the phase interface position.

This inconsistency is still under investigation.
Possible explanations include too idealized boundary conditions in the simulation and a bad insulation regarding the experiment.
These must be investigated in the future, either by extending the simulation or by conducting tailored experiments using a redesigned experimental setup with less uncertainties than introduced by the present setup.

In general, both our capability to simulate complex multi-physics problems, as well as our capability to acquire data grew extensively in recent years. Optimal combination of both that result in high quality model validation strategies at high 
spatio-temporal resolution are, however, rare. Standard practice is often rather to compare sophisticated models to a sparse data set, or to analyze large data sets with very idealized models. Exceptions exist for certain processes, e.g. as relevant for meteorology, but cannot be easily extended to arbitrary process models. 
On our way to explore sophisticated model validation strategies at high spatio-temporal resolution, we proposed to set up a tailored laboratory experiment and designed data processing to match ideally with our major simulation goal. Inconsistencies between the simulation and experiment are accessible, which would be hard to acknowledge if validation had been done with sparse data only. Our next steps will be two-fold, namely specifically investigating the inconsistencies between model and experiments in the concrete conducted validation study, and more generally continue to work on flexible, integrated validation strategies for coupled multi-physics systems.

\section*{Acknowledgement}
The project was funded in part by the Excellence Initiative of the German Federal and State Governments.
It is supported by the Federal Ministry for Economic Affairs and Energy, Germany, on the basis of a decision by the German Bundestag (FKZ: 50 NA 1502). It is part of the Enceladus Explorer initiative of the DLR Space Administration.

\bibliography{schueller2018arxiv}{}
\bibliographystyle{plain}

\ifx\justbeingincluded\undefined
\end{document}
\fi